\documentstyle[12pt]{article}
\newcommand{\be}{\begin{equation}}
\newcommand{\ee}{\end{equation}}
\newcommand{\ben}{\begin{eqnarray}
\displaystyle}
\newcommand{\een}{\end{eqnarray}}

\newcommand{\la}{{\lambda}}

\newcommand{\cK}{{\cal K}}
\newcommand{\cD}{{\cal D}}
\newcommand{\cM}{{\cal M}}
\newcommand{\cJ}{{\cal J}}

\newcommand{\cO}{{\cal O}}

\newcommand{\cL}{{\cal L}}

\newcommand{\p}{\partial}
\newcommand{\na}{\nabla}
\newcommand{\hna}{\hat \nabla}

\newcommand{\epk}{{(\ep^{I})^\dagger}}

\newcommand{\th}{{\tilde h}}
\newcommand{\tna}{{\tilde \nabla}}
\newcommand{\tga}{{\tilde \gamma}}
\newcommand{\tep}{{\tilde \epsilon}}
\newcommand{\tS}{{\tilde \Sigma}}
\newcommand{\hA}{\hat A}

\newcommand{\ha}{\hat a}
\newcommand{\hb}{\hat b}
\newcommand{\hi}{\hat i}
\newcommand{\hj}{\hat j}
\newcommand{\hc}{\hat c}
\newcommand{\hk}{\hat k}
\newcommand{\ho}{\hat 0}
\newcommand{\hp}{\hat 5}
\newcommand{\hje}{\hat 1}
\newcommand{\hd}{\hat 2}
\newcommand{\httt}{\hat 3}

\newcommand{\ep}{\epsilon}

\newcommand{\ga}{\gamma}

\begin{document}

\title{Positive Mass Theorem for Black Holes in Einstein-Maxwell
Axion-dilaton Gravity}

\author{Marek Rogatko \\
Technical University of Lublin \\
20-618 Lublin, Nadbystrzycka 40, Poland \\
rogat@tytan.umcs.lublin.pl \\
rogat@akropolis.pol.lublin.pl}

\date{\today}

\maketitle
\smallskip
{\bf PACS numbers:} 04.50.+h, 97.60.L.\\
\begin{abstract}
We presented the proof of the positive mass theorem for black holes
in Einstein-Maxwell axion-dilaton gravity being the low-energy limit of
the heterotic string theory. We show that the total
mass of a spacetime containg a black hole is greater
or equal to
the square root of the sum of
squares of the adequate {\it dilaton-electric} and
{\it dilaton-magnetic} charges.
\end{abstract}


\baselineskip=18pt
\section{Introduction}
In general relativity, global quantities as a total energy or total
angular momentum of an isolated system do not arise naturaly as in
special relativity. There were heavy attempts to prove that the mass of
an isolated system is positive. 
The story began in 1959 when partial results were obtained by Araki
\cite{ar} 
and Brill \cite{br}. The first complete proof of 
the positive energy theorem
was deviced by Schoen and Yau \cite{sch}. 
Shortly afterwards, Witten \cite{wi} conceived the elegant proof that the ADM
mass of an asymptotically flat spacetime containing matter satisfying
the dominant energy conditions is non-negative and vanishes in the case
of a flat spacetime. His reasoning was based on the analysis of spinors
fulfilling a Dirac-like equation on a three-dimensional spacelike
hypersurface. Soon after, Parker and Taubes \cite{pa} gave a mathematically
rigorous proof of the positive energy theorem. They completed and
simplified the original arguments presented in \cite{wi}.
\par
Nester's attitude \cite{ne} treated the problem of the positive energy mass
theorem in a fully covariant way, in order to avoid technical difficulties
concerning the three-dimensional truncation of the four-dimensional
divergence theorem. 
In several papers \cite{bondi} the issue of extending Witten's proof of the
positivity of energy at spatial infinity to a proof in the case of null
infinity was considered. The authors established the positivity of the
Bondi mass.
Recently Penrose {\it et al.} \cite{pe} based on a causal structure and
focusing proved the positive mass theorem.
\par
Soon after, the positive energy theorem was established, 
similar techniques were used to prove several extensions of this result.
Gibbons and Hull \cite{gi} proved 
the positive mass theorem for Einstein-Maxwell
theory and were able to derive a lower bound for the mass of the spacetime.
The positive mass theorem was also generalized to hold in
supergravity \cite{sup} and in Kaluza-Klein theory \cite{kk,lee}. 
In \cite{sup1} it
was shown that an
inclusion of Yang-Mills fields, Yang-Mills-Higgs, dilaton interactions
implied that self-gravitating solitons saturated a gravitational version
of the Bogomolnyi bound on energy. In paper \cite{ro} the 
proof of the positive
mass theorem in the case of the low-energy string theory, i.e, the so-called
Einstein-Maxwell axion-dilaton gravity, was presented. A lower bound
for the mass of a spacetime was derived in the theory under consideration.
\par
Gibbons {\it et al.} \cite{mass}
extended the positive mass theorem to
asymptotically flat manifolds containing black holes. Recently,
Herzlich \cite{he} provided 
the rigorous mathematical proof of the black holes positive
mass theorem. In \cite{cur}
Gibbons extended
the Geroch-Wald-Jang-Hu\-is\-ken-Il\-ma\-nen 
approach to the positive energy
problem and gave a negative 
lower bound for the mass of asymptotically Anti-de-Sitter spacetimes
containing horizons. It was also shown that the method gave a lower
bound for the mass of
time-symmetric initial data sets for black holes with scalar and vectors.
For a review of the positive energy theorem story see, e.g.,\cite{co}.
\par
In this paper we want to provide some continuity with the work 
\cite{ro} and to some extent generalize it. We shall consider
the problem of the black hole
positive mass theorem in the so-called
Einstein-Maxwell axion-dilaton gravity
being the so-called the low-energy limit of the heterotic string theory.
Our paper is organized as follows. In Sec.II 
the basic field equations of the theory under consideration were
presented and the non-negativity of the black hole mass was established.
Sec.III summarizes our results.
\par
In our paper
the metric $g_{\mu \nu}$ has signature $(- +++)$ and $\ga$ matrices
obey the standard condition $\{ \ga^{\mu}, \ga^{\nu} \} = 2g^{\mu \nu}$,
$\sigma^{\mu \nu} = {1 \over 4}[ \ga^{\mu}, \ga^{\nu} ]$.
Greek indices change from 0 to 3, while Latin
ones from 1 to 3. Indices with hats are refered to an orthonormal frame
in which $\ga^{\ha}$ matrices are hermitian and $\ga^{\ho}$ antihermitian.


\section{Black Holes Positive Mass Theorem}
The so-called low-energy limit of superstring theories provides
an interesting generalization of the Einstein-Maxwell (EM) theory. A
simplified model of this kind in an Einstein-Maxwell axion-dilaton
(EMAD) 
coupled system containing a metric $g_{\mu \nu}$, $U(1)$ vector fields
$A_{\mu}$, a dilaton $\phi$ and three-index antisymmetric tensor field
$H_{\alpha \beta \ga}$. The action has the form \cite{wil}
\be
I = \int d^4 x \sqrt{-g} 
\left [ R - 2(\na \phi)^{2} - {1 \over 3} H_{\alpha \beta \ga}
H^{\alpha \beta \ga} -
e^{-2\phi} F_{\alpha \beta} F^{\alpha \beta} \right ] + I_{matter},
\label{ac}
\ee
where the strength of the gauge fields are descibed by
$F_{\mu \nu} = 2\na_{[\mu} A_{\nu]}$ and
the three index antisymmetric tensor is defined by
\be
H_{\alpha \beta \ga} = \na_{\alpha}B_{\beta \ga} - A_{\alpha}F_{\beta
\ga} + cyclic
\ee
In four dimensions $H_{\alpha \beta \ga}$ is equivalent to the
Peccei-Quin pseudoscalar and may be written as follows:
\be
H_{\alpha \beta \ga} = {1 \over 2}\ep_{\alpha \beta \ga \delta}
e^{4 \phi} \na^{\delta} a.
\label{def}
\ee
As a consequence of the definition (\ref{def}) the action (\ref{ac})
yields
\be
I = \int d^4 x \sqrt{-g} 
\left [	R - 2(\na \phi)^{2} - e^{-2\phi} 
F_{\alpha \beta} F^{\alpha \beta} - a F_{\mu \nu} \ast F_{\mu \nu}
\right ] + I_{matter},
\label{act}
\ee
where  $\ast F_{\mu \nu} = 
{1 \over 2}\ep_{\mu \nu \delta \rho} F^{\delta \rho}$ .
The resulting equations of motion, derived from the variational
principle become
\ben
\na_{\mu} \left ( e^{-2\phi} F^{\mu \nu} + a \ast F^{\mu \nu}
\right ) = \cJ^{\nu}(matter), \\
\na_{\mu} \left ( \ast F^{\mu \nu} \right ) = 0, \\
\na_{\mu} \na^{\mu} \phi - {1 \over 2}e^{4\phi}\na_{\mu} a \na^{\mu} a
+ {1 \over 2} e^{-2\phi} F^2 = 0, \\
\na_{\mu} \na^{\mu} a + 4 \na_{\ga} \phi \na^{\ga} a - e^{-4 \phi}
F_{\mu \nu} \ast F^{\mu \nu} = 0, \\
G_{\mu \nu} = T_{\mu \nu}(matter) + T_{\mu \nu}(F, \phi, a),
\een
where $ T_{\mu \nu}(F, \phi, a)
 = {- 2 \delta I \over \sqrt{-g} \delta g^{\mu \nu}}$
the energy momentum tensor for $U(1)$ gauge fields, axion and
dilaton fields reads
\ben
T_{\mu \nu}(F, \phi, a) = e^{2 \phi} \left (
4 F_{\mu \rho} F_{\nu}{}{}^{\rho} - g_{\mu \nu} F^2 \right )
&-& g_{\mu \nu} \left [
2 (\na \phi )^2 + {1 \over 2} e^{4 \phi} ( \na a )^2 \right ] \\
\nonumber
&+& \na_{\mu} \phi \na_{\nu} \phi + e^{4 \phi}\na_{\mu} a \na_{\nu} a.
\een
In what follows
we will consider a spacelike hypersurface $\Sigma$, with induced
metric $h_{ij}$ imbeded in four-dimensional spacetime
$(\cM, g_{\mu \nu})$. 
A spacelike hypersurface $\Sigma$ is assumed to be asymptotically flat,
i.e., there exists a flat metric $\delta_{ij}$ defined outside a
compact set $\cal G$ such that $\Sigma - \cal G$ is diffeomorphic to
the complement of a compact set in $R^3$, ~$h_{ij} = \delta_{ij} 
+ \cO \left ( {1 \over r} \right )$, ~$ K_{ab} = 
\cO \left ( {1 \over r^2} \right )$. $K_{ab}$ is an extrinsic curvature
of $\Sigma$.
The hypersurface element is denoted by $dA^{i}$ while the boundary at
spatial infinity by $\p \Sigma$.
One
chooses the tetrad so that the zero indices vector
is orthogonal to the hypersurface $\Sigma$.
The manifold under consideration
will contain a black hole and the main task will be to evaluate the mass
of the black hole.
\par
In order to do so we introduce
the supercovariant derivative acting on a spinor field $\ep_{I}$
is given by the formula \cite{ka,ro}
\be
\hna^{(4)}_{\mu}\ep_{I} = \na^{(4)}_{\mu}{}{}{} 
\ep_{I} + {i \over 2}e^{2 \phi}
\na^{(4)}_{\mu}a ~\ep_{I}
+ {i \over 2}e^{- \phi} F_{\alpha \beta}
\ga^{\alpha} \ga^{\beta} \ga_{\mu} \alpha_{IK} \ep^{K},
\label{sup}
\ee
where $I, K$ stand for $SO(4)$ indices, $\alpha_{IK}$ is $SO(4)$ matrix
\cite{cr}.
The suprecovariant derivative (\ref{sup}) can be thought as a
supersymmetry transformation about non-trivial gravitational, scalar
and $U(1)$ gauge backgrounds \cite{sup,ro}.\\
Projecting  the four-dimensional
supercovariant derivative into the hypersurface $\Sigma$ and
multiplying the result by $\ga^{\ha}$ and in the end
equating the outcome to zero, one gets the Witten equation given by
\be
\ga^{\ha} \na^{(3)}_{\ha}{}{} \ep_{I} + {1 \over 2} K \ga^{\ho} \ep_{I}
+ {i \over 2} e^{2 \phi} \ga^{\hb} \na^{(3)}_{\hb} a \enskip \ep_{I}
- {i \over 4} e^{- \phi} \ga^{\hb} \left (
E_{\ha} - 2 B_{\ha} \ga_{\hp} \right )
\ga^{\ho} \ga^{\ha} \ga_{\hb} \alpha_{IK} \ep^{K} = 0.
\label{www}
\ee
where $K = K_{\ha}{}{}^{\ha}$ is the triad component of the second
fundamental form of the hypersurface $\Sigma$,
and $\ga^{\hp} = \ga^{\ho}\ga^{\hje}\ga^{\hd}\ga^{\httt}$,
$\left ( \ga_{\hp} \right )^2 = - 1$. The adequate components of
$F_{\mu \nu}$ have the forms
\be
F_{\hb \ho} = E_{\hb}, \qquad
F_{\ha \hb} = \ep_{\ha \hb \hc}B^{\hc}.
\ee
As in the derivation of the positive mass theorem \cite{ro}, it will be
convenient to define the quantity defined as follows:
\be
\delta \la_{I} = \ga^{\alpha} \na^{(4)}_{\alpha} \phi \enskip
 \ep_{I}
+ {i \over 2} e^{2 \phi} \ga^{\beta} \na^{(4)}_{\beta} a \enskip \ep_{I}
- {i \over 8}e^{- \phi} F_{\alpha \beta}
\ga^{\alpha} \ga^{\beta} \alpha_{IK} \ep^{K}.
\label{la}
\ee
The motivation for introducing $\delta \la_{I}$
is to achieve the desired mass bound for black holes. However, the
expression (\ref{la}) has also the motivation as the supersymmetry
transformation laws of the appropriate particles in the associated
supergravity model.
\par
Taking into account equations of motion, after lengthy calculations
we reached to the indentity
\ben 
\label{sur}
\int_{H} dA^{i} ~
\epk \hna^{(4)}_{i} \ep_{I} + 
\int_{\Sigma} d\Sigma ~
(\hna^{(4) i}~ \ep^{I})^{\dagger} \hna^{(4)}_{i} ~\ep_{I} &+& \\ \nonumber
{1 \over 2} \int_{\Sigma} d\Sigma ~\epk \left [
T_{\ho \ho}(matter) + T_{\ho \ha}(matter) \ga^{\ho} \ga^{\ha} \right ]
\ep_{I} &-& i \int_{\Sigma} d\Sigma ~
e^{-\phi} \epk \ga^{\ho} \left ( 
\cJ_{\ho}{}{}^{E} - \ga^{\hp} \cJ_{\ho}{}{}^M \right ) \alpha_{IK}~
\ep^{K}  \\ \nonumber
+  \int_{\Sigma} d\Sigma ~
(\delta \la^{I})^{\dagger}
\delta \la_{I} +
\int_{\Sigma} d\Sigma ~
\epk \cK \alpha_{IK}~ \ep^{K}
&=& \int_{S_{\infty}} dA^{i} ~
\epk \hna^{(4)}_{i} \ep_{I},
\een
where $\cJ_{\mu}^{E}$ is the electric current while $\cJ_{\mu}^{M}$
is the magnetic current. The
energy momentum tensor of the matter fields is equal to
\be
T_{\mu \nu}(matter) = T_{\mu \nu}(total) - T_{\mu \nu}(F, \phi, a),
\ee
while $\cK$ denotes
\ben
\cK &=& e^{-\phi} \left (
3 E^{\hi} \na^{(3)}_{\hi} \phi \ga^{\ho} - \ep^{\hi \hj \hk}
\na^{(3)}_{\hi} E_{\hj}
\ga_{\hk} \ga_{\hp} + 2 \ep^{\hi \hj \hk} \na^{(3)}_{\hi} a~ B_{\hj}
\ga_{\hk} \right ) \\ \nonumber
&+& {e^{\phi} \over 4} \left (
B^{\hi} \na^{(3)}_{\hi} a \ga^{\ho} \ga^{\hp} - \ep_{\hi \hj \hk} B^{\hi}
\na^{(3) \hj} a~ \ga^{\hk} \right ).
\een
Now we turn to the question of 
a spacelike hypersurface $\Sigma$ with an
boundary $H$. We shall work
in the tetrad which $e^{\rightarrow}_{\ho}$ unit
zero-vector is orthogonal to $\Sigma$
and unit one-vector $e^{\rightarrow}_{\hje}$
is orthogonal to $H$. The remainning ones lie in $H$.
It happened that one cannot require the condition of vanishing
$\ep_{I}$ on $H$ to be satisfied. As was discussed in \cite{mass}, the
Witten's equation 
would imply that the derivative of $\ep_{I}$ transversal to
$H$ also disappeared. So that one has
$\ep_{I} = 0$ everywhere. Then, we need to restrict the freedom of
$\ep_{I}$ on $H$. Following \cite{mass}, one 
chooses the boundary conditions
as follows:
\be
\ga^{\hje} \ga^{\ho} \ep_{I} - \ep_{I} = 0.
\label{b}
\ee
The above condition restricts the freedom of
spinors $\ep_{I}$ on $H$
by half. It is caused so by the fact that the matrix
$\ga^{\hje} \ga^{\ho}$ has eigenvalues $\pm 1$, with eigenspaces of
dimensions equal to two. Taking into account the Witten's equation,
$\ga^{\ha}\hna^{(4)}_{\ha} \ep_{I} = 0$,
admitting 
a solution satisfying the asymptotic requirements and the boundary
conditions (\ref{b}) we will have a closer look at the surface term on
$H$ in equation (\ref{sur}). After some 
algebraic manipulations, the surface
term on $H$ can be written as
\ben \label{h}
\int_{H}dA^{\hc} ~
\epk \hna^{(4)}_{\hc} \ep_{I} = 
- {1 \over 2} \int_{H}dA ~ \epk \left [
\left ( K + J - K_{\hje \hje} \right ) \ga^{\hje} \ga^{\ho} \ep_{I}
+ 2 \ga^{\hje} \ga^{\hA} \cD_{\hA} \ep_{I} \right ] + \\ \nonumber 
- {i \over 8} \int_{H} dA ~ e^{- \phi} \epk
\left [ 2 \ga^{\ho} \left ( E_{\hje} - \ga_{\hp} B_{\hje} \right )
\right ] \alpha_{IK}~ \ep^{K}
- {i \over 2} \int_{H} dA ~ e^{2 \phi} \epk
\ga^{\hje} \ga^{\hA}
\na^{(2)}_{\hA} a~
\ep_{I}. 
\een
where we set $\hA = \hd, \httt$ and $J$ denotes the mean curvature of
$H$ in $\Sigma$. By $\cD_{\hA} \ep_{I}$ we defined \cite{mass}
the following derivative:
\be
\cD_{\hA} \ep_{I} = \na^{(2)}_{\hA} \ep_{I} + {1 \over 2}
K_{\hA \hje} \ga^{\hje} \ga^{\ho} \ep_{I}.
\ee
An inspection of
equation (\ref{b}) easily shows that
the matrix $\ga^{\hje} \ga^{\ho}$ anticommutes with the operator
$\cL = \ga^{\hje} \ga^{\hA} \cD_{\hA}$, thus the second term on the
right-hand side of equation (\ref{h}) vanishes if $\ep_{I}$ satisfies
the boundary conditions. Since then also $\epk = - \epk 
\ga^{\hje} \ga^{\ho}$ and $\epk \cL \ep_{I} = - \epk \cL \ep_{I}$,
which implies that $\epk \cL \ep_{I} = 0$. Applying the same arguments, one
finds that matrices $\ga^{\ho}$, $\ga^{\ho} \ga_{\hp}$ anticommute with
$\ga^{\hje} \ga^{\ho}$ then the third term of expression (\ref{h})
disappear.\\
Moreover, if $H$ is a future apparent horizon \cite{haw}, 
one has that $K + J - K_{\hje \hje} = 0$.
If $H$ is a past apparent horizon the same situation takes place
\cite{mass}. Then, one uses boundary conditions like (\ref{b}) but
with a minus sign.\\
The last term of the considered equation vanishes if we impose the
additional condition for axion fields, namely that there is no {\it
axion currents} on the horizon surface, i.e., $\na^{(2)}_{\hA} a = 0$.\\
Thus we see that the right-hand side of
equation (\ref{h}) vanishes.\\
Further, we shall assume that the matter energy momentum tensor obeys the 
following energy condition
\be
T_{\ho \ho}(matter) \ge \left [
T_{\ho j}(matter) T_{\ho}{}{}^{j}(matter) + \left ( \cJ_{\ho}^{E} 
\right )^{2} + \left ( \cJ_{\ho}^{M} \right )^{2} \right ]^{1 \over 2},
\label{c1}
\ee
and moreover, we impose the additional conditions, as follows:
\be
\delta \la_{I} = 0, \qquad \cK = 0.
\label{c2}
\ee
The first relation is motivated by the invariance of the entire system
under the supersymmetry transformation, while the other inputs
relations among the fields appearring in the theory under
consideration. The very similar condition was obtained in
the energy bounds studies in Einstein-Maxwell-dilaton system \cite{sup1}
and in EMAD system without interior boundaries \cite{ro}.
\par
Now we would like to remark on the existence of the solutions of
equation (\ref{www}). Following the reasoning presented in \cite{mass},
instead of dealing with the spinor fields which tend asymptotically to
constant values at large distances on $\Sigma$ we take into account a
conformal transformation compactify the hypersurface by adding a point
at infinity. The metric $\th_{ij}$ on a compact manifold $\tS$
will be conformally related to the metric on $\Sigma$, $\th_{ij} =
\Omega^4 h_{ij}$. 
The conformal factor $\Omega$ is required to compactify an
asymptotically flat hypersurface $\Sigma$ and it is of the form
$\Omega = {1 \over r} + \cO \left ( {1 \over r^2} \right )$,
while
$K = \cO \left ( {1 \over r^2} \right )$.
As we shall take into account the behaviour on the
horizon we set $P_{a} = 0 $ \cite{mass}. On the above smooth manifold
with the boudary $H$ the spinor fields $\tep_{I} =
{\ep_{I} \over \Omega^2}$
obey the relation as follows:
\be
\tga^{\ha} \tna^{(3)}_{\ha}{}{} \tep_{I} + {1 \over 2} 
\Omega^2 K \ga^{\ho} \tep_{I}
+ {i \over 2} \Omega^2
e^{2 \phi} \ga^{\hb} \na^{(3)}_{\hb} a~ \tep_{I}
- {i \over 4} \Omega ^2 e^{- \phi} \ga^{\hb} \left (
E_{\ha} - 2 B_{\ha} \ga_{\hp} \right )
\ga^{\ho} \ga^{\ha} \ga_{\hb} \alpha_{IK} \tep^{K} = 0,
\label{cw}
\ee
where $\tga^{b} = \Omega^{-2} \ga^{b}$ and $\tna_{i}$ is the covariant
derivative with respect to $\th_{ab}$.
Suppose that equation (\ref{cw}) has a non-zero solution $\tep_{I(1)}$
obeying the boundary conditions (\ref{b}). Then $\Omega^2 \tep_{I(1)}$
would be a spinor field on $\Sigma$ which satisfies the Witten
equation, the boundary conditions on $H$ and which decreases to zero at
infinity like ${1 \over r^2}$. In the light of equation (\ref{h})
and conditions (\ref{c1}) and (\ref{c2}), it
will be a contradiction because of vanishing of the boundary terms. It
shows that
the conformal Witten's equation with the boundary (\ref{b})
has no zero modes on $\tS$. All these arguments suggest 
that there exist
a Green's function to this elliptic boundary value problem \cite{hor}.
In this way, one can obtain solutions to the Witten's equation on
$\Sigma$ fulfilling the boundary conditions (\ref{b}) and approaching a
constant spinor at infinity. \\
Probably, the existence of a non-zero solution to equation (\ref{www})
can be rigourously proved by means of an isomorphism between some
adapted Sobolev spaces. We hope to return to this question elsewhere.
\par
To complete our considerations concerning equation (\ref{sur})
we shall consider 
the surface term at infinity.
The relationship between the integral of the surface term at infinity
and the ADM mass may be demonstrated by considering solutions of the
Witten equation. By virtue of the direct generalization of the 
arguments given in
\cite{wi}, one gets
\be
\int_{S_{\infty}} dA^{i} ~
\epk \hna^{(4)}_{i} \ep_{I} = {1 \over 2}
(\ep^{I}_{\infty})^{\dagger} M \ep_{I \infty} -
{i \over 8} (\ep^{I}_{\infty})^{\dagger} \ga^{\ho} \left (
Q_{(F-\phi)} - P_{(F-\phi)} \right ) \alpha_{IK} \ep^{K}_{\infty},
\ee
where $\ep_{I \infty}$ is a spinor field which is constant in some
chart around infinity. $M$ is the ADM
mass defined by $M =
\sqrt{P_{ADM}P_{ADM}}$, where $P_{ADM}$ is the ADM four momentum.  
The {\it dilaton-electric} and {\it dilaton-magnetic} charges 
of the black hole are defined respectively as
\be
Q_{(F-\phi)} = \int_{S_{\infty}} dS^{i} e^{- \phi_{\infty}} E_{i},
\qquad
P_{(F-\phi)} = \int_{S_{\infty}} dS^{i} e^{- \phi_{\infty}} B_{i}.
\ee
The non-negativity of the left-hand side of equation (\ref{sur}) for any 
$\ep_{I \infty}$, implies that
\be
M \ge \sqrt{Q_{(F-\phi)}^2 + P_{(F-\phi)}^{2}}.
\ee
We have thus proved the positive black hole mass theorem 
in EMAD gravity.

\section{Conclusions}
In summary, we have studied the positive mass theorem for
black holes in EMAD gravity being the low-energy limit of the heterotic
string theory. We analysed an asymptotically flat spacelike hypersurface
with induced metric $h_{ij}$, containing black holes. We have
considered
spinors obeying the Dirac type equation on this hypersurface. Using the
classical Witten's arguments we show that in four-dimensional manifold
satisfying the energy condition (\ref{c1})
and the two other requirements (\ref{c2})
imposed on fields in the theory, 
the ADM mass of the black hole in the theory under consideration
is nonnegative, provided that the square of the total mass of a
spacetime containing the black hole is greater or equal to the sum of
squares of the total {\it dilaton electric} and {\it dilaton magnetic}
charges of the black hole.

\vspace{2cm}
{\bf Acknowledgement} \\
I would like to thank the unknown referees for very useful comments.

\eject

\end{document}